\documentstyle[epsfig,aps,floats,prc,preprint]{revtex}
\newcommand\pt {$p_t$\ }

\newcommand\pim {\pi^-}
\newcommand\pip {\pi^+}

\begin{document}
\draft
\title{ Proton and Pion Production in Au+Au Collisions at 10.8A GeV/c}

\author{
  J.~Barrette$^5$, R.~Bellwied$^{9}$, 
  S.~Bennett$^{9}$, R.~Bersch$^7$, P.~Braun-Munzinger$^2$, 
  W.~C.~Chang$^7$, W.~E.~Cleland$^6$, M.~Clemen$^6$, 
  J.~Cole$^4$, T.~M.~Cormier$^{9}$, 
  Y.~Dai$^5$, G.~David$^1$, J.~Dee$^7$, O.~Dietzsch$^8$, M.~Drigert$^4$,
  K.~Filimonov$^3$, S.~C.~Johnson$^7$, 
  J.~R.~Hall$^9$, T.~K.~Hemmick$^7$, N.~Herrmann$^2$, B.~Hong$^2$, 
  Y.~Kwon$^7$,
  R.~Lacasse$^5$, Q.~Li$^{9}$, T.~W.~Ludlam$^1$,
  S.~K.~Mark$^5$, R.~Matheus$^{9}$, S.~McCorkle$^1$, J.~T.~Murgatroyd$^9$,
  D.~Mi\'{s}kowiec$^2$,
  E.~O'Brien$^1$,  
  S.~Panitkin$^7$, P.~Paul$^7$, T.~Piazza$^7$, M.~Pollack$^7$, 
  C.~Pruneau$^9$, Y. ~J. ~Qi$^5$,
  M.~N.~Rao$^7$, E.~Reber$^4$, M.~Rosati$^5$, 
  N.~C.~daSilva$^8$, S.~Sedykh$^7$, U.~Sonnadara$^6$, J.~Stachel$^3$, 
  E.~M.~Takagui$^8$, V. ~Topor ~Pop$^5$, S.~Voloshin$^6$, 
  T.~B.~Vongpaseuth$^7$, G.~Wang$^5$, J.~P.~Wessels$^3$, 
  C.~L.~Woody$^1$, N.~Xu$^7$,
  Y.~Zhang$^7$, C.~Zou$^7$\\ 
(E877 Collaboration)
}
\address{
 $^1$ Brookhaven National Laboratory, Upton, NY 11973\\
 $^2$ Gesellschaft f\"ur Schwerionenforschung, 64291 Darmstadt, Germany\\
 $^3$ Universit\"at Heidelberg, 69120 Heidelberg, Germany\\
 $^4$ Idaho National Engineering Laboratory, Idaho Falls, ID 83402\\
 $^5$ McGill University, Montreal, Canada\\
 $^6$ University of Pittsburgh, Pittsburgh, PA 15260\\
 $^7$ SUNY, Stony Brook, NY 11794\\
 $^8$ University of S\~ao Paulo, Brazil\\
 $^9$ Wayne State University, Detroit, MI 48202\\
}

\date{April 26, 2000}
\maketitle

\begin{abstract} We present proton and pion 
tranverse momentum spectra and rapidity distributions for
Au+Au collisions at 10.8A GeV/c. 
The proton
spectra exhibit collective transverse flow effects.
Evidence of the influence of the Coulomb
interaction from the  fireball is found in the pion transverse
momentum spectra.
The data are compared with the predictions of the RQMD event
generator. 
\end{abstract}

\section{INTRODUCTION}

     One of the main motivations of the relativistic heavy-ion
program is the formation of baryonic matter at extreme temperatures
and densities, and the subsequent characterization of such matter.
Of central importance
is the ability to understand to what extent the nuclear matter has been
compressed and heated.  
At the AGS, Au+Au collisions at about 11A GeV/c have 
been characterized rather completely in terms of the global observables
such as transverse energy \cite{auauet} and charged particle multiplicity 
\cite{charged}. From these measurements, it was concluded that 
central Au+Au collisions exhibit 
a large degree of stopping and leads to larger energy density than 
lighter systems at the same bombarding energy.
Through comparisons with models that reproduce the experimental data, 
it has been concluded that one reaches 
baryon densities of up to
10 times normal nuclear matter density and energy densities of the order of 
2 GeV/fm$^3$ in the center of the fireball \cite{johana99}. 
These are in the range of parameters where based on QCD calculations one 
expects a baryon-rich deconfined phase. 
Distributions in rapidity and transverse 
momentum of the particles emitted in such collisions contain more 
complete information about the degree of stopping attained and provide
more sensitive tests of the validity of the model predictions.
Semi-inclusive proton and pion distributions from central
Au+Au reactions at 11.6A GeV/c have been reported in 
Ref \cite{e802ahle}. It was observed that the rapidity distribution
for protons has a maximum around mid-rapidity consistent with a large degree
of stopping and hence high baryon density.
In this paper we present
measurements of the forward-rapidity
inclusive double differential multiplicities
and rapidity distributions for proton and pions in central 
Au+Au collisions at a beam momentum of 10.8A GeV/c.  

\section{EXPERIMENT}

The experiment was performed using the E877 apparatus at the AGS at 
the Brookhaven National Laboratory.  The E877 set-up used in this
experiment is
an upgrade of the E814 apparatus which has been previously
described in Ref. \cite{e814}. The device features nearly $4\pi$ calorimetry
surrounding the target.
The information from the calorimeters was used in this work primarly
for centrality selection.
Charged particles emitted in the forward
direction ($ -134\,$ mrad $ < \theta_{horizontal} < 16\, $ mrad 
and $-11\,$ mrad $< \theta_{vertical} < 11\,$ mrad) 
are analyzed by a magnetic spectrometer.
The spectrometer has been significantly upgraded so as to handle the
high multiplicity of charged particles produced in the Au collisions.
Inserted between two high resolution drift chambers are four new multiwire
proportional chambers which aid in the pattern
recognition by confirming links between the drift chambers.
The section of the
spectrometer which receives primary beam particles was purposely
made insensitive to avoid saturation and failure of the tracking devices.   
A new 150 slat high-resolution
time-of-flight hodoscope \cite{roger98}
was installed downstream of the tracking
chambers, 12 meters from the target.  The average time
resolution of the hodoscope was 85 ps.  
Particles are identified by measuring their momentum and
velocity.  The system provides proton-pion
separation, at the 2.5 $\sigma$ level, up to a momentum of 8.8 GeV/c.
The momentum resolution ($\delta$p/p $\sim$ 3\%) of the tracking system is
dominated by multiple scattering.
Contamination from high momentum kaons is estimated
to contribute less than 5 \% to the measured proton yield 
at all rapidities and to the experimental $\pi^+$ yield 
at rapidities $y < 3.8$.
The kaon contamination will increase slowly with rapidity is an estimated
to reach up to 10 \% of the measured pion yield at $y$ = 4.4.

In order to reconstruct the pion and proton spectra, the data 
must to be corrected for the spectrometer acceptance and the effects
of the various conditions introduced in the analysis. The acceptance 
corrections for the distributions have been calculated using 
Monte Carlo simulation. The acceptance corrections on the final data
sample were calculated as function of rapidity y and transverse
momentum $p_t$ by propagating generated particles through the 
E877 apparatus. All known effects of the spectrometer geometry ,
detector resolutions, kinematics and cuts were included. 
For pion spectra the influence of pion decay is 
included in the program used to calculate the 
acceptance of the spectrometer.

For central Au+Au collisions the mean spectrometer occupancy 
is about 7-8
charged particles per event, and is strongly position dependent in the
bending plane of the spectrometer, varying from 1\% to 18\% per 
drift chamber sense wire.  Track
reconstruction efficiency decreases abruptly
when two tracks are closer than twice the
wire spacing in a drift chamber. The occupancy
variation introduces a position dependent efficiency for
track reconstruction and is a source of systematic
error.  Two correction procedures were developed as part of a
set of independent data analyses.  
Each of the procedures used the {\em measured} occupancy in all
tracking devices as its basis, thereby avoiding model dependence in
the correction. 
The first method \cite{lacasse}
involved a detailed model of the track reconstruction efficiency as a
function of track separation. Each track is given a weight that,
to first order,
accounts for the tracks that are lost due to the presence of the track seen. 
No tracks are removed from the data sample in this method. 
The other method rejected every pair of tracks which in a given detector
failed a minimum separation cut below which 
the tracking efficiency is less than unity \cite{piazza,voloshinnote1}. 
In this case the effect of the cut was calculated either
by adding virtual tracks to the events  
and computing 
the probability that such tracks fail the minimum cut.
The correction factor varies from 1.1 to 1.3, the largest 
correction being for tracks passing near the deadened beam region.
The two correction methods give very similar results with 
maximum differences of 5\%.  The second correction
procedure was also tested in a simulation using the GEANT\cite{geant}
package.  
Application of the correction procedure to 
the Monte Carlo generated data tracked
through the spectrometer reproduces the initial
distribution with maximum deviations of 5\%.  
Considering the agreement both between the two correction
methods and with the GEANT simulation, and the systematic uncertainties in
overall single track reconstruction efficiency, we deduce a combined
systematic error of less than 10\%.
All data points  in the figures have statistical error bars only,
which are often smaller than the data point symbol.

Central collisions were defined as those producing high transverse
energy in the pseudorapidity interval $ -0.5 < \eta < 0.8$,
where $\eta = -\rm{ln[\tan(\theta/2)]}$,
such that $\sigma_{central}$ is 4\% of the geometric cross
section.
The resolution on the centrality has been estimated by studying 
the fluctuation in the transverse energy ($E_t$) distribution in the
solid angle covered by the target calorimeter TCAL that is used to
determine the collision centrality. This results in a smooth 
cut-off in the impact parameter distribution that is taken into 
account in generating the calculated spectra.
Based on RQMD simulations it is estimated that the selected events
correspond to a mean impact parameter of about 2.4 fm. 
For the $0-4 \%$ centrality cut this resolution in impact parameter is
estimated to be $\approx$ 0.7 fm.

\section{RESULTS}

\subsection{Protons}

\begin{figure}[hbt!]
\centering
\epsfig{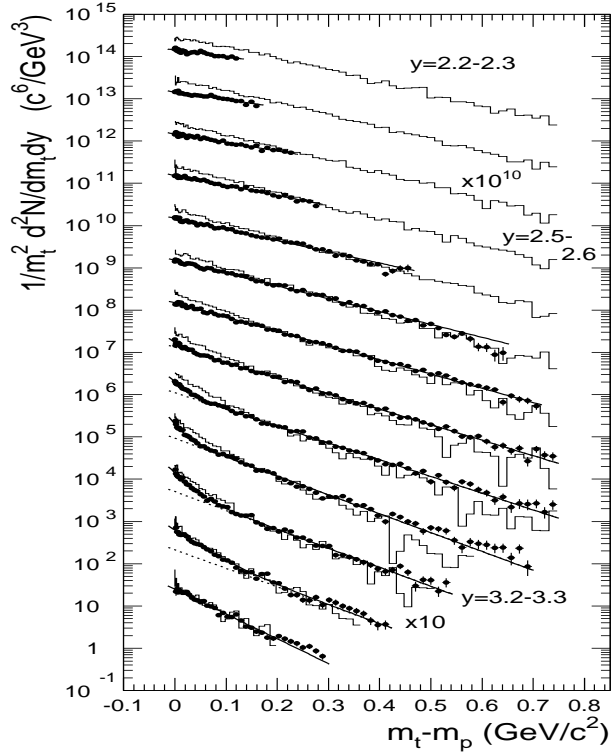} 
\caption{Transverse mass spectra for protons produced in central
($\sigma_{central}$/$\sigma_{geom}$=0.04)
Au+Au collisions. The dots correspond to 
constant $p_{t}$ bins of 20 MeV, and $y$ bins of
0.1 units.  Beginning with rapidity bin $y$=3.3-3.4,  
spectra have been
multiplied by successively increasing factors of ten.  
Full lines are two component exponential fits to the data. The histograms 
are RQMD predictions. }
\label{Fig:mt}
\end{figure}

Fig.~\ref{Fig:mt} shows the
measured proton transverse mass spectra for central Au+Au collisions.
The vertical axis is ($1/m_t^2) \times (d^2N/dm_tdy)$, the representation
in which a Boltzmann (or thermal) distribution is a pure exponential
in $m_t$ ($m_t = \sqrt{p_t^2+m^2}$).  
The spectra are close to exponential. One notes, however, a steep component
at low $m_t$ for rapidities near $y_{beam}$=3.14, 
which in Si+A collisions was identified as due to the
projectile spectator nucleons\cite{frag}.

Also plotted are the results of two component
Boltzmann fits to the data (full lines).  
The dashed lines near beam rapidity show the contribution from the component
with large inverse slope parameter  
T$_B$ that can be attributed to emission from the central
\lq\lq fireball\rq\rq. This component always dominates at $m_t-m$ 
larger than 0.1 GeV/c$^2$.
The histograms correspond to the prediction of the RQMD 
event generator \cite{rqmd1} (RQMD 1.08).


As it was done for the experimental data, the analysis of the RQMD 
generated events included selection of 
the 4\% most central collisions  
from the produced $E_{t}$
in the pseudorapidity interval $ -0.5 < \eta < 0.8$. 
RQMD treats the projectile and target as an unbound ensemble of
nucleons with Fermi energy distributions. This introduces a large unphysical
excess in the number of calculated spectator nucleons. 
These were removed by considering
in the calculated spectra only protons that have had one or more
interactions.

RQMD reproduces quite well the average multiplicities but the 
calculated spectra are consistently steeper than the data.

\begin{figure}[h!]
\centering
\epsfig{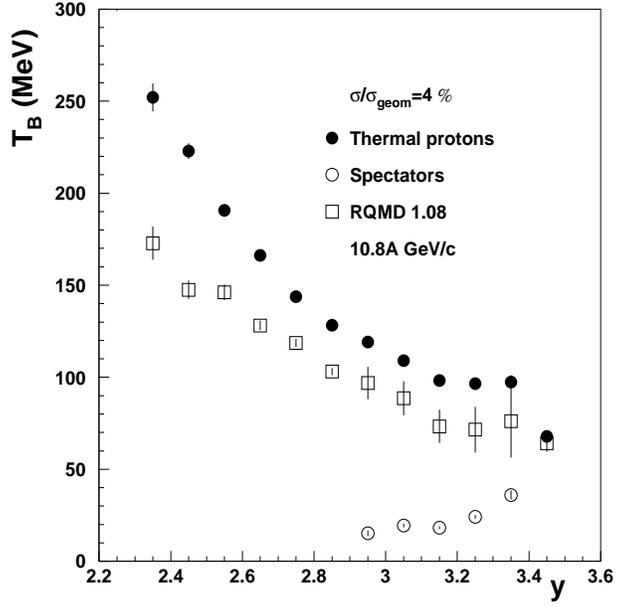} 
\vskip 0.5cm
\caption{Inverse slope parameters deduced
from fits to  the proton $m_t$ spectra for central
Au+Au collisions. The solid circles correspond to the high $m_t$ 
component and the open circles correspond to the 
low $m_t$ component. The squares 
are the results for the high $m_t$
component of a similar fit to the 
calculated spectra.}
\label{Fig:temp}
\end{figure}

The systematic differences between the data and the model predictions
are better shown in 
Fig.~\ref{Fig:temp} where we compare the values of fitted 
inverse slope parameters T$_B$ as a function of rapidity,
with RQMD predictions. At each rapidity the calculated spectra were 
fitted over the range covered by experimental data.

The rapidity dependences of the measured and predicted
inverse slope parameters are very similar 
but the RQMD spectra give values that are systematically
about 20\% lower than the data. 
The derived slope parameters increase
very rapidly near central rapidity. This is different than
what was observed
for Si+Al collisions\cite{e814,e802} 
and also deviate from the 1/cosh($y$) dependence 
expected for an isotropically expanding fireball.


The observed behavior can be explained by the presence
of strong transverse collective flow 
in the measured spectra near midrapidity. 
Transverse collective flow leads to deviation from the
exponential shape expected for a pure thermal source 
\cite{peter95,mattiello,peter96}.

In particular, transverse
flow results in flatter distribution (i.e. higher value of T$_B$)
at low $m_t$. 
This flattening which has been observed \cite{e802ahle}
is predicted to increase with the flow velocity
and with the mass of the produced particles.
The fact that the measured spectra are flatter 
than the calculated spectra
suggests that the pure cascade version of RQMD as implemented in RQMD 1.08   
predicts a too small collective flow.
A similar conclusion has been reached in Ref. \cite{protflow} 
from a study of the proton and pion azimuthal distributions measured 
relative to the reaction plane.

\begin{figure}[hbt!]
\centering
\epsfig{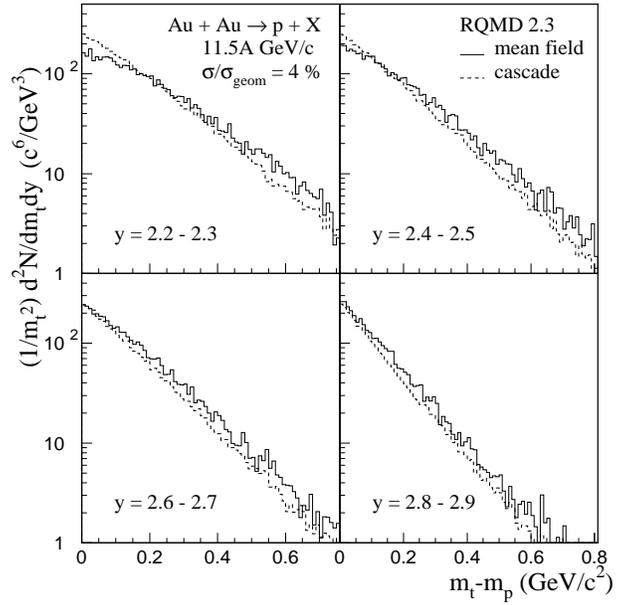} 
\vskip 0.5cm
\caption{Comparison of the proton
transverse mass spectra for central Au+Au collisons at 11.5A Gev/c 
predicted by RQMD 2.3 run in a cascade mode (dashed histograms) 
and including the effects of mean field (full histograms).}
\label{Fig:rqmd2.3}
\end{figure}

This is also consistent with the results
shown in Fig.~\ref{Fig:rqmd2.3} where we compare the proton 
spectra predicted by RQMD 2.3 for Au+Au collisions 
in cascade mode and including the effect of mean field.
The introduction of mean-field in RQMD gives somewhat
flatter spectra particularly at low value of $m_t$ 
and close to mid-rapidity.

\begin{figure}[hbt!]
\centering
\epsfig{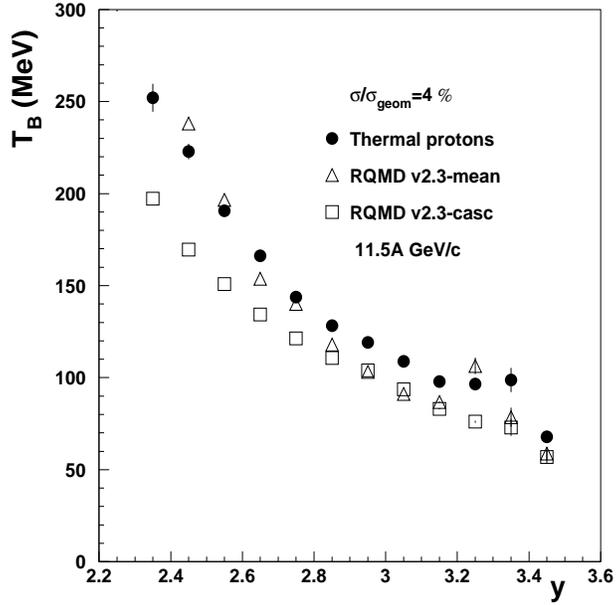} 
\vskip 0.5cm
\caption{Inverse slope parameters deduced
from fits to  the proton $m_t$ spectra for central
Au+Au collisions. The solid circles correspond to fits to the high $m_t$ 
component of the experimental spectra at 10.8 GeV/c. The open symbol 
are the results of a similar fit to the 
calculated spectra at 11.5 GeV/c predicted by RQMD 2.3 - in cascade
mode(squares) and with mean field included (triangles).}
\label{Fig:tb_rqmd23}
\end{figure}

The predicted effect of the mean field on the deduced inverse
slope parameter in our acceptance is shown in Fig. \ref{Fig:tb_rqmd23}.
The cascade results are in good agreement with those in
Fig. ~\ref{Fig:temp} when taken into account the shift 
$\Delta y = 0.03$ in the center of mass rapidity between the two
energies. The introduction of mean field in RQMD 2.3 has little 
effect on the inverse slope parameter at the most forward rapidities ,
but give results in much better agreement with the data closer to midrapidity.



\begin{figure}[hbt!]
\centering
\epsfig{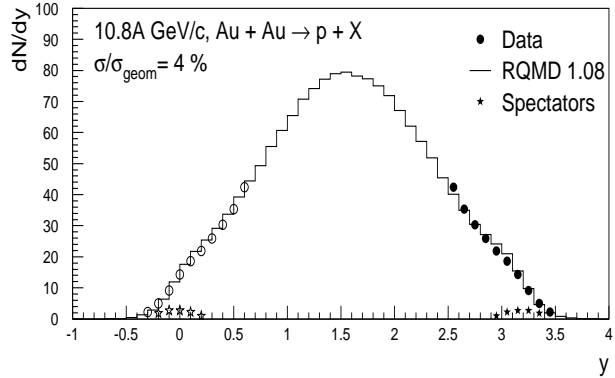} 
\vskip 0.5cm
\caption{Rapidity distributions for protons in central Au+Au
collisions.  
Filled symbols represent measured data, open symbols
reflected data. The stars  
correspond to the contribution from the
low $m_t$ component to the total measured multiplicity (circles).  
The histogram shows  the results
from RQMD calculations.}
\label{Fig:dndy}
\end{figure}

The proton rapidity distribution $dN/dy$ is presented in Fig.~\ref{Fig:dndy}.
It has been obtained  by integrating over
the transverse mass using the measurements
where available and extending the integral analytically to $p_t=0$ and
$p_t\rightarrow\infty$ using the results from the exponential 
fits.
This procedure was tested using the  spectra calculated with RQMD. 
The data points nearest to the center of mass have been omitted
since the limited m$_t$ range of the measured spectra 
results in large systematic errors.
It is estimated that the extrapolation
leads to an overestimation of the yield for the first data point
in Fig.~\ref{Fig:dndy}  at $y$=2.55 by a maximum of roughly 15 \% while the
systematic error is negligible for data points at $y \geq 2.85$.
The stars represent the contribution to the total proton
multiplicity from the low $m_t$ component
calculated from the fitted spectra. As expected, it is centered near
$y_{beam}$=3.14. It is a relatively small contribution to the total proton
multiplicity. The peak near beam rapidity integrates to one proton.
In the measured rapidity range, RQMD (histogram) 
reproduces fairly well the measured
rapidity distribution.  
The calculated distribution shows a weak inflexion near beam rapidity 
that can be associated with spectator-like protons that have had
minimal interaction in the collision.


\subsection{Pions}
 
The transverse mass spectra for  $\pi^+$ and $\pi^-$ are presented in
Fig.~\ref{Fig:pionmt}. 
The solid curves show 
Boltzmann fits to the data. 
Exponential functions are fitted to the data 
above to $m_t - m_{\pi}$ $\geq$ 175 MeV/c$^2$
in the rapidity bins where available,
$y \leq 3.4$ for $\pi^+$ and  $y \leq 3.9$ for $\pi^-$,and starting 
from $m_t - m_{\pi}$ $=$ 0 GeV/c$^2$ at higher rapidities.

As observed for Si+Pb system 
\cite{delta}, the spectra show a significant enhancement over a 
Boltzmann distribution at low $m_t$. This has been attributed to the
contribution from the decay of $\Delta$ resonances and higher mass
resonances \cite{delta,brown,bal91,sollfrank90}. 
The rapidity dependence of the shape of the spectra and  
the magnitude of low \pt enhancement attributed to resonance decay
pions are well reproduced by RQMD.
The calculation slightly over-estimates the yield of negative
pions especially at high rapidities. 

\begin{figure}[hbt!]
\centering
\epsfig{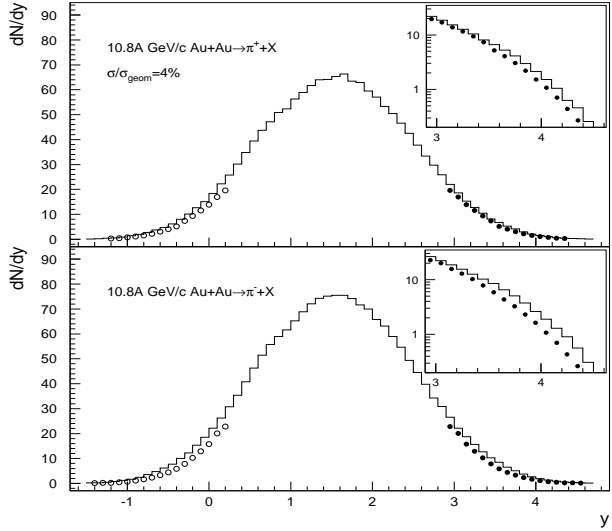} 
\vskip 0.5cm
\caption{Rapidity distributions for charged pions (dots)
in central Au+Au collisions. Measurements are
reflected (open symbols) about mid rapidity. Also shown are the results from
RQMD calculations (histogram).
The insets show the high rapidity part of the distributions on a logarithmic 
scale.}
\label{Fig:pidndy}
\end{figure}

The pion rapidity distributions for $\pi^+$ and $\pi^-$
are obtained by integrating over the transverse mass,
using the measured yield where available and extrapolating the 
integral analytically to large $m_t$. The results are  
compared to the predictions of RQMD in Fig.~\ref{Fig:pidndy}.
The shape of the rapidity distribution is
rather well reproduced by RQMD over the entire range in rapidity
covered by the present data. 
A similar agreement was also observed for the Si+A 
systems~\cite{qm933,e8028}.
Small but significant 
deviations in the tail of the distribution are 
better visualized in the insets of the figure  where the same
distributions are shown on a logarithmic 
scale. In the range of our measurement
the yields of both the positive and negative pions are systematically 
overpredicted by the RQMD~1.08 model with deviations
that increase with rapidity and reaches a factor $\simeq$ 1.5 at 
rapidity y=4.


\subsection{$\pi^{-}$/$\pi^{+}$ ratio}

\begin{figure}[hbt!]
\centering
\epsfig{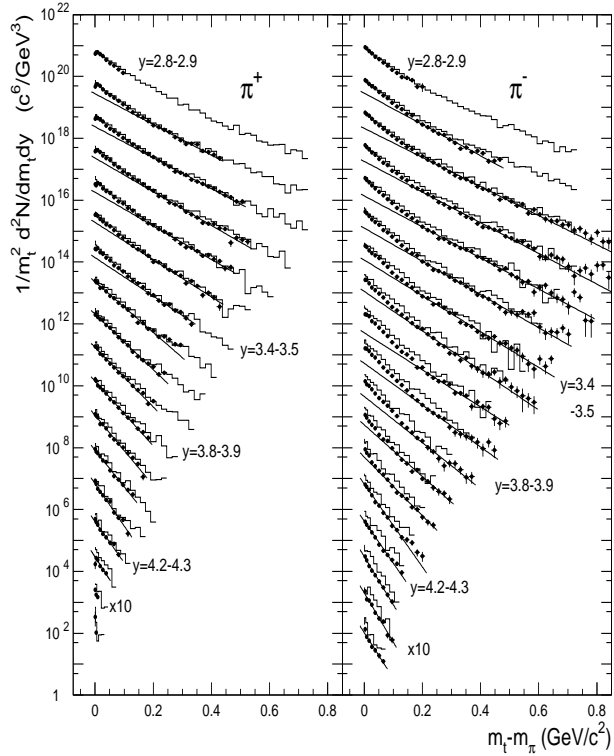}
\vskip 0.5cm
\caption{Pion transverse mass spectra for central 
($\sigma_{central}$/$\sigma_{geom}$=0.04) 
Au+Au collisions. Beginning with y=4.4-4.5, the spectra have 
been multiplied by successively increasing factors of ten.    
Full lines are the results of Boltzmann fits to the data
over a limited range (see text). 
The histograms are RQMD predictions.}
\label{Fig:pionmt}
\end{figure}

\begin{figure}[t!]
\centering
\epsfig{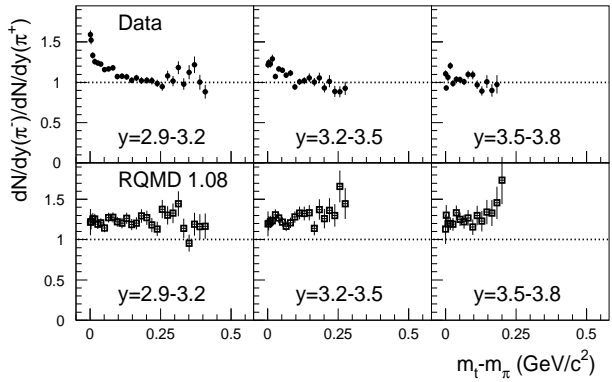}  
\vskip 0.5cm
\caption{Top row; transverse mass dependence of
the experimental pion cross section ratio $\pi^-$/$\pi^+$
for three rapidity intervals. Bottom row; 
corresponding RQMD predictions for the $\pi^-$/$\pi^+$ cross section
ratio.}
\label{Fig:piratio}
\end{figure}

Close inspection of the transverse mass curves of
Fig.~\ref{Fig:pionmt} reveals that the deviations from a pure 
exponential emission are systematically smaller for
$\pi^{+}$ than for $\pi^{-}$.   
This charge asymmetry of the pion distributions is better studied by
plotting  the ratio of $\pi^{-}$ to $\pi^{+}$ cross section as done 
in Fig.~\ref{Fig:piratio}.  
In order to improve statistical errors 
on the ratio, the data were grouped into three larger rapidity bins.
A strong charge asymmetry is observed starting at 
$m_t-m_{\pi}< 0.2$~GeV/c$^2$ with a maximum value of
$dN/dy(\pi^-) / dN/dy(\pi^+)\simeq 1.6$ 
at $m_t-m_{\pi}=0$~GeV/c$^2$ for the
$y=2.9-3.2$ rapidity slice. 
The measured asymmetry systematically decreases 
as a function of rapidity.
The observed rapidity dependence is in agreement with the  
charge asymmetry 
measured by the E802 collaboration near 
$y_{\rm cm}$ for similar centrality~\cite{e802ahle}. 

The observed pion charge asymmetry and its rapidity dependence 
can be attributed to the different Coulomb
potentials seen by the two type of pions at freeze-out.
The Coulomb origin of this effect is supported by the results of the 
RQMD calculations which do not take into account final state
interaction of the reaction products
(second row of Fig.~\ref{Fig:piratio}). As expected, RQMD predicts 
no significant $m_t$ dependence in the ratio of the pion cross section
with an overall excess of negative pions of $\approx$ 20\%.

In heavy systems,
anomalously large $\pim/\pip$ ratios 
have been attributed to the effect of Coulomb interaction 
over a wide range of beam energies
~\cite{bene79,bert80,wolf79,naga81,wolf82,bal95,na4496,wagner1}.
At the AGS, the first evidence of Coulomb effects in heavy systems was
reported in $^{28}$Si induced reactions \cite{qm937,e86695}.



We have used a simple model to study if 
the measured rapidity and m$_t$ dependence of the 
$\pi^-/\pi^+$ ratio can be explained by the Coulomb interaction.
The model is built along the line of the argument 
presented in~\cite{gyua81,qm964} and similar 
to that discussed in \cite{wagner1}. 
It uses an effective central Coulomb potential 
to simplify the difficult many-body problem.

Assuming that the {\it fireball} is at rest in the center-of-mass
and neglecting its time
evolution, the effective Coulomb potential $V_C$ seen by a 
singly-charged test particle is given by,
\begin{equation}
V_C = \frac{Z_{\rm eff}\cdot{e^2}}{r_i} 
\label{eq:cpot}
\end{equation}
where the effective charge $Z_{\rm eff}<Z_A+Z_B$ is expected to be
smaller than the total charge present in $A+B$ collisions and 
$r_i$ is the radius at which the particle leaves the fireball.
After exiting from the Coulomb potential well,
the measured energy ${E(p)}$ of a test particle of positive/negative charge
becomes 
\begin{equation}
E({p})=E({p_i})\pm V_C
\end{equation}



where $p_i$ are the initial momentum and position at freeze-out.
The number of particles per unit of momentum is then expressed as
\begin{equation} 
n (\mbox{$p$}) 
= n(\mbox{$p_i$})\frac{d^3p_i}{d^3p} 
= \frac{p_i E(p_i)}{p E(p)} n(\mbox{$p_i$})
\end{equation}
where $n(p_i)$ is the initial particle distribution and 
${d^3p_i}/{d^3p}$ is the 
Jacobian deduced from energy conservation~\cite{qm964}.
The initial single particle distribution $n(\mbox{$p_i$})$
is therefore changed in magnitude as well as distorted by the Coulomb field. 
The Coulomb field induced distortion ${\mathcal C}$,
can then be written as :
\begin{eqnarray}
{\mathcal C}^\pm & = & \frac{p_i E(p_i)}{p E(p)}\\
                 & = & \sqrt{p^2 \mp 2 E(p) V_C + V_C^2} \cdot {\frac{(E(p) \mp V_C)}{p E(p)}}
\end{eqnarray} 
for positively/negatively charged particles.
The Coulomb distortion of the pion ratio is expressed in 
terms of measured momenta $p$ as
\begin{equation}
\frac{\pi^-}{\pi^+}(p) = {\rm R}\cdot\frac{{\mathcal
C}^-}{{\mathcal C}^+}\cdot{\frac{n(p^-_i)}{n(p^+_i)}}\\
\end{equation}
\begin{equation}
 = {\rm
R}\cdot\frac{\sqrt{p^2+2 E(p) V_C + V_C^2}}{\sqrt{p^2-2 E(p) V_C+V_C^2}}
\cdot{\frac{(E(p) + V_C)}{(E(p) - V_C)}}\cdot{\frac{n(p^-_i)}{n(p^+_i)}}
\end{equation}
where ${\rm R}$ is an overall normalization constant.
The transverse momentum and rapidity dependences are introduced by
substituting $E = {\rm m_t} \cosh(y)$ and $p=\sqrt{E^2-m^2}$.
Thus, the model predicts the shape of the $\pi^-/\pi^+$ ratio 
as well as its rapidity dependence with only two free parameters: the 
effective Coulomb potential $V_C$ and the normalization constant $\rm R$. 

The model neglects the deflection of the particle trajectories
induced by the Coulomb interaction. 
It is assumed that the initial angle of the particle ($p_t/p_z$)
is conserved.
Possible quantum mechanical effects are ignored and 
we also neglect the possible contribution from the spectators.
Such a contribution has been discussed at lower 
energies~\cite{gyua81,libb79}.
However, our analysis concerns only the most central collisions
where the fraction of the total charge carried by the spectators 
is relatively small. 
Also, the projectile spectators are far from the central fireball at
t$\approx$9 fm/c, the typical pion emission time determined by 2 particle
correlation studies \cite{e814xu}.
The model also assumes a static effective potential that is identical 
for all the particles. 
This approximation is not valid in general.
However, here we discuss mainly high rapidity pions that are in
the tail of the $dN/dy$ rapidity distribution.
Thus, on the average, these pions see a relatively 
similar central fireball at freeze out.
This also justifies the neglect of 
the expansion of the fireball.
These approximations are less valid 
as one approaches mid-rapidity.

The undistorted $n(p^\pm_i)$ distributions are not 
accessible experimentally.
In the present calculation, in agreement with the RQMD prediction,
we assume that both distributions are similar in shape.
The initial spectral shapes are generated using mean values of the
parameters used in the two-exponential description of the 
data (see Fig.~\ref{Fig:pionmt}).

\begin{figure}[h!]
\centering
\epsfig{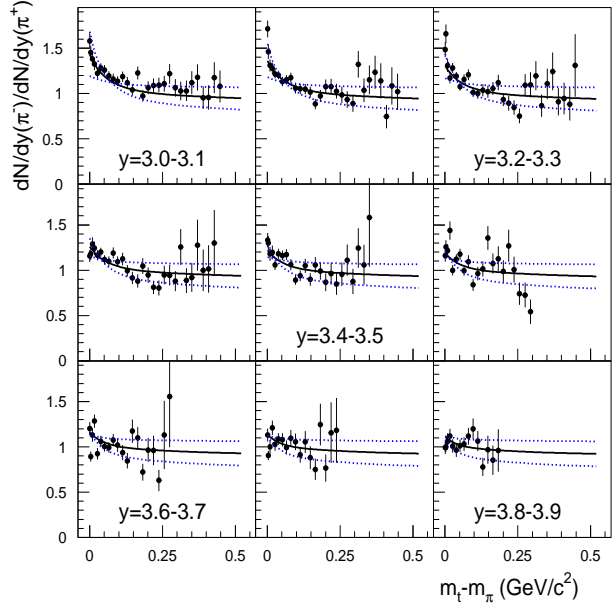} 
\vskip 0.5cm 
\caption{The $\pi^- / \pi^+$ ratio as a function of
transverse mass for nine rapidity intervals.
The solid line is the result from a Coulomb distortion calculation.
The dotted lines show the statistical error of the fit.}
\label{fig:pi_ratios_err}
\end{figure}

The model predictions are compared to the measured ratios 
of pion transverse mass spectra in Fig.~\ref{fig:pi_ratios_err}.
The predictions of the Coulomb distortion model (solid lines) 
are obtained from a fit to the first four measured distributions
where the observed effect is the largest.
The values of $V_C\approx 31\pm 22$~MeV and ${\rm R} \approx 0.92\pm 0.14$ 
are found to best describe the shape and rapidity dependences
of the pion ratio.
These parameters are found to be still in agreement with the data 
at larger rapidities. 
The errors of the fitted values (dotted lines) are rather large, because 
there is a strong correlation between the effect
of the relative normalization constant $R$ and that of the 
effective Coulomb potential value.  

The Coulomb induced distortion has 
two components: a Coulomb factor 
${\mathcal C}^- / {\mathcal C}^+$ and an amplitude 
factor ${n(p^-_i)}/{n(p^+_i)}$ that is related to 
the shape of the initial pion spectra.
The effects of R and $V_C$ can not be distinguished
above transverse mass values of about 0.25~GeV/c$^2$ since the
Coulomb distortion is rather flat and affects mainly 
the ratio of the pion yield.
At low $m_t$,
the two factors distort the shape of the pion ratio in a similar way, 
but quantitatively 
most of the distortion is attributable to the Coulomb factor.

Note, that secondary $\pim$ from lambda decay will generate an excess
of negative pions. This effect was studied by  a Monte Carlo
simulation using as input the lambda spectrum predicted by RQMD and
taking into account the acceptance of the E877 spectrometer. 
This calculation shows that in the rapidity range discussed here the
contribution from $\Lambda$ decay is negligible above 
$m_t-m_{\pim}$ = 0.1 GeV/c$^2$ and contribute at most 
 10-15\% of the observed asymmetry at  $m_t-m_{\pim}$ = 0,
where this effect is maximum.We verified the influence of this 
uncertainty on the derived values of $V_C$ by modifying the data 
set in agreement with the calculated influence of $\Lambda$ decay.
The effect on the results was found to be of the order of 2 MeV and
thus negligible when compare to the large uncertainty 
originating from the strong correlation between $V_C$ and $R$.

The present value of $V_C$ is in a good agreement with
the Coulomb potential obtained
from a similar analysis 
of the pion spectra in Au+Au collisions at
a bombarding energy of 1 GeV/nucleon \cite{wagner1}.


The uncertainty on the value of the $\rm R$ and $V_C$ constants 
could be improved with a larger data sample, 
\emph{e.g.} by providing a measurement that would extend to larger 
values of $m_t$.
Such data 
would better fix the relative values of $\rm R$ and $V_C$ 
while the shape at low $m_t$ would fix the magnitude of $V_C$.
We remark also, that a more precise estimate of the effect
of the Coulomb field on the measured pion spectra should take into
account the effect of radial flow and emission time 
\cite{mos95,barz97}.

\begin{figure}[hbt!]
\centering
\epsfig{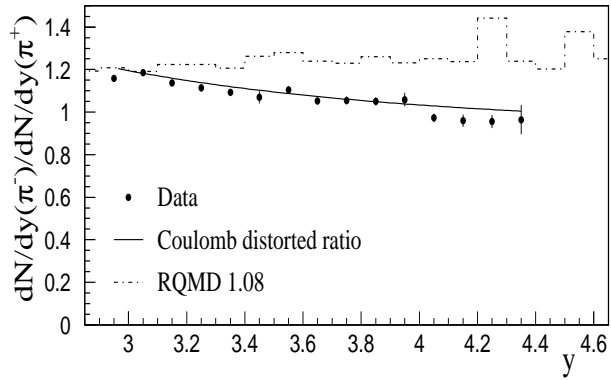} 
\vskip 0.5cm
\caption{Rapidity distribution of the 
experimental $\pi^-$/$\pi^+$ cross section ratio (dots). 
Also shown are  
the calculated ratio taking into account the effect of Coulomb
distortion (full line) and the results from RQMD calculations (histogram).}
\label{Fig:pidnratio}
\end{figure}

The pion rapidity distribution should also show effects of the
Coulomb interaction since this interaction not only affects the
transverse momentum but also the final rapidity of the emitted 
pions.
We have shown in Fig.~\ref{Fig:pidndy}  that the $\pi^+$ 
and $\pi^-$ have similar rapidity distributions 
reasonably well reproduced by RQMD with, however, increasing
deviations at very forward rapidity. 
In Fig.~\ref{Fig:pidnratio}, the ratio of the pion yield 
$dN/dy(\pi^-) / dN/dy(\pi^+)$ is plotted as a function 
of rapidity.
One observes that the ratio shows
a systematic decrease towards high rapidities while
the RQMD~1.08 model 
(dashed histogram) predicts 
a very flat if anything  slightly opposite rapidity dependence.

We have used the Coulomb distortion model developed here to
determine if we can provide a consistent description of the
effects observed both in the transverse mass spectra and in the rapidity
$dN/dy$ distribution.
The $dN/dy$ ratios calculated using the parameters
$V_C$ = 31 MeV and $R$ = 0.92   deduced from the shape of 
the $m_t$ spectra are shown
by the solid line in Fig.~\ref{Fig:pidnratio}.
The calculated ratios are obtained by
integrating the calculated distorted
transverse mass distribution.


The rapidity dependence of the data is very well reproduced 
by the model over the entire measured rapidity range.
The effective Coulomb field in Au+Au collisions is strong enough 
to significantly distort the rapidity distributions of light particles
and, as shown in the figure,
the ratio of pion $dN/dy$ can vary 
by more than 20\% depending on the rapidity at which it is measured.
This result shows that light particle yield ratios 
should be interpreted with care
and the comparison of yield ratios from experiments performed  
at different rapidities should take into account 
potential Coulomb effects.

It is interesting to compare the expected
Coulomb potential viewed by the first pions that are ejected 
from the fireball to the effective Coulomb potential $V_C$ 
obtained from our simple model. 
Using a simple geometrical model it is
estimated that the mean total charge
of the participants for the present event sample is $Z \approx 150$.
Handbury-Brown-Twiss (HBT) analysis of pion correlations
provides an estimate of the fireball radius.
One-dimensional HBT analysis of Au+Au central
gives radius   of $\approx$ 6.0 fm ~\cite{pipi877,qm966,baker96,qm967}.
These numbers lead to uniformly-charged-sphere Coulomb 
potential values of $V \approx 36$ MeV,
 an estimate  consisten with the
value obtained from our simple model of the
Coulomb interaction.

\section{CONCLUSIONS}

In summary, we have presented new results on the 
spectra and multiplicity 
distributions of protons and pions emitted at
forward rapidity in Au+Au collisions at 10.8A GeV/c.
When combined with the data from E802 experiment \cite{e802ahle}
we now have a measurements of proton and pion production in 
central Au+Au collisions at AGS energy over the complete phase space.

The shape of the rapidity distribution of the protons 
over the entire range in rapidity covered by the present data,
is  rather well reproduced by RQMD 1.08 . 
The measured transverse momentum spectra
have, however, larger inverse slope parameters 
than predicted by the model. This is consistent
with the presence
of a larger collective tranverse flow than predicted by a pure cascade model. 

The shape of the spectra and mutiplicity distributions of the pions
are very well described by RQMD. Systematic differences between the $\pi^-$
and $\pi^+$ spectra at low values of the tranverse momentum 
are, however, 
observed that are not predicted by the model. It is shown using a simple
model that the observed effect is consistent with that expected from
the different Coulomb potential felt by the two types of pions.
Our simple analysis shows
that the $\pi^-/\pi^+$ ratio provides an additional way, 
complementary to the Hanbury-Brown-Twiss interferometry and
particle correlation analysis, 
for studying source sizes and dynamics.

\section*{ACKNOWLEDGMENTS}

We wish to thank the Brookhaven Tandem and AGS staff for their excellent
support and are particularly grateful for the expert help of W.
McGahern and Dr. H. Brown. We also wish to acknowledge the important 
technical support provided by R. Hutter and J. ~Sondericker.  
Financial support by the US DoE, the NSF,
the Canadian NSERC, and CNPq Brazil is acknowledged.

\newpage 

\end{document}